\documentclass{PoS}
\let\OLDthebibliography\thebibliography
\renewcommand\thebibliography[1]{
  \OLDthebibliography{#1}
  \setlength{\parskip}{-0.5pt}
  \setlength{\itemsep}{-0.5pt plus 0.ex}
}

\title{Chandra spectroscopy of Rapid Burster type-I X-ray bursts}

\ShortTitle{Chandra spectroscopy of Rapid Burster type-I X-ray bursts}

\author{\speaker{J.J.M. in 't Zand} \& T. Bagnoli\\
        SRON Netherlands Institute for Space Research, Utrecht,
        the Netherlands\\
        E-mail: \email{jeanz@sron.nl, tullio.bagnoli@gmail.com}}

\author{C. D'Angelo \& A. Patruno\\
        Leiden University, the Netherlands\\
        E-mail: \email{caroline.dangelo@gmail.com; patruno@strw.leidenuniv.nl}}

\author{D.K. Galloway\\
  School of Physics \& Astronomy and Monash Centre for Astrophysics,
  Monash University, Clayton, Australia\\
        E-mail: \email{duncan.galloway@monash.edu}}

\author{M.B.M. van der Klis, A.L. Watts\\
        Anton Pannekoek Institute, University of Amsterdam, the Netherlands\\
        E-mail: \email{m.b.m.vanderklis@uva.nl; a.l.watts@uva.nl}}

\author{H.L. Marshall\\
        Massachusetts Institute of Technology, Cambridge, MA 02139, USA\\
        E-mail: \email{hermanm@space.mit.edu}}

\abstract{We observed the Rapid Burster with Chandra when it was in
  the 'banana' state that usually precedes the type-II X-ray bursting
  'island' state for which the source is particularly known. We
  employed the High-Energy Transmission Grating Spectrometer in
  combination with the ACIS-S detector in continuous clocking
  mode. The observation yielded 20 thermonuclear type-I X-ray bursts
  emitted from the neutron star surface with recurrence times between
  0.9 and 1.2 hr, and an e-folding decay time scale of 1 min. We
  searched for narrow spectral features in the burst emission that
  could constrain the composition of the ashes of the nuclear burning
  and the compactness of the neutron star, but found none. The upper
  limit on the equivalent width of narrow absorption lines between 2
  and 6 keV is between 5 and 20 eV (single trial 3$\sigma$ confidence
  level) and on those of absorption edges between 150 and 400 eV. The
  latter numbers are comparable to the levels predicted by Weinberg,
  Bildsten \& Schatz \cite{wei06} for Eddington-limited thermonuclear
  bursts.}

\FullConference{11th INTEGRAL Conference Gamma-Ray Astrophysics in Multi-Wavelength Perspective,\\
		10-14 October 2016\\
		Amsterdam, The Netherlands}

\begin{document}

\section{Introduction}
\label{intro}

An important question in (astro)physics concerns the behavior of
matter at supranuclear densities and low temperatures such as in
neutron stars. Due to the mathematical difficulty in describing
multi-particle interactions under such circumstances, there is no
unambiguous prediction and one needs to resort to observation and
experiment. Macroscopically, one describes the behavior of matter by
the equation of state (EOS). That, on its turn, prescribes the
relationship between the mass and radius of neutron stars. Therefore,
if one were able to measure mass and radius of a set of neutron stars,
one would be able to infer the sought-after EOS (e.g.,
\cite{lattimer2007,watts2016}). This is one of the fundamental
pursuits in astrophysics. One of the roads where this pursuit takes
place \cite{paerels2009} is that of type-I X-ray bursts. These are
thermonuclear shell flashes on H/He-accreting neutron stars in
low-mass X-ray binaries (LMXBs). Such a flash results for $\sim1$ min
in a luminous signal from the neutron star in 1-10 keV X-rays. If one
were in the position to detect identifiable narrow spectral features
in the flash emission, one could obtain, through the gravitational
redshift, constraints on both mass and radius. The detection
probability is improved if the neutron star is slowly spinning, thus
avoiding Doppler smearing of narrow spectral features.

The Rapid Burster, or MXB 1730-335 \cite{lewin1976,lewin1993}, is a
puzzling LMXB. It earned its nickname because of its unique exhibition
of bright quickly repetitive so-called type-II X-ray bursts that have
different time profiles and spectra than type-I bursts and are thought
to be due to an accretion instability. The Rapid Burster is a
transient X-ray source in the globular cluster Liller 1 (distance
8.1$pm1.0$ kpc; $N_{\rm H}=(1.8\pm0.1)\times10^{22}$~cm$^{-2}$;
\cite{saracino2015}) that goes into outburst about every half a year
for about one month \cite{masetti2002,simon2008}. It is the only LMXB
that exhibits both type-II and type-I X-ray bursts.

Bagnoli et al. \cite{bag13,bag14,bag15a,bag15b} made a comprehensive
study of the large Rapid Burster data set (total observation time
close to 2.4 Ms) provided by observation with the high-throughput
Proportional Counter Array on the Rossi X-ray Timing Explorer during
its 16 years of operation. They found the following features (see also
\cite{guerriero1999}): 1) it is an Atoll source (for a definition, see
\cite{hasinger1989}); 2) there is a clear division in behavior between
the two Atoll states: the island and banana state; 3) it exhibits the
quickly repetitive type-II X-ray bursts in the island state; 4) it
acts as an ordinary LMXB without type-II bursts in the banana state;
5) it exhibits peculiar behavior in the transition between the two
states with strong few-Hz quasi-periodic oscillations, dipped type-I
X-ray bursts and very long (up to 10$^3$ s) type-II bursts; 6) it
sometimes exhibits types of variability ('heartbeat' and 'theta') that
is otherwise only seen in the peculiar black-hole LMXBs GRS 1915+105
and IGR J17091-3624; 7) the behavior of the type-I bursts suggests
that the neutron star in MXB 1730-335 is slowly spinning, with a spin
period of maybe about 1 s \cite{bag13}, in contrast to most other
LMXBs where neutron stars spin at hundreds of Hz \cite{patruno2012}.

\begin{figure}[t]
\begin{center}
\includegraphics[width=\columnwidth,angle=0,trim=0cm 1cm 0cm
  14cm]{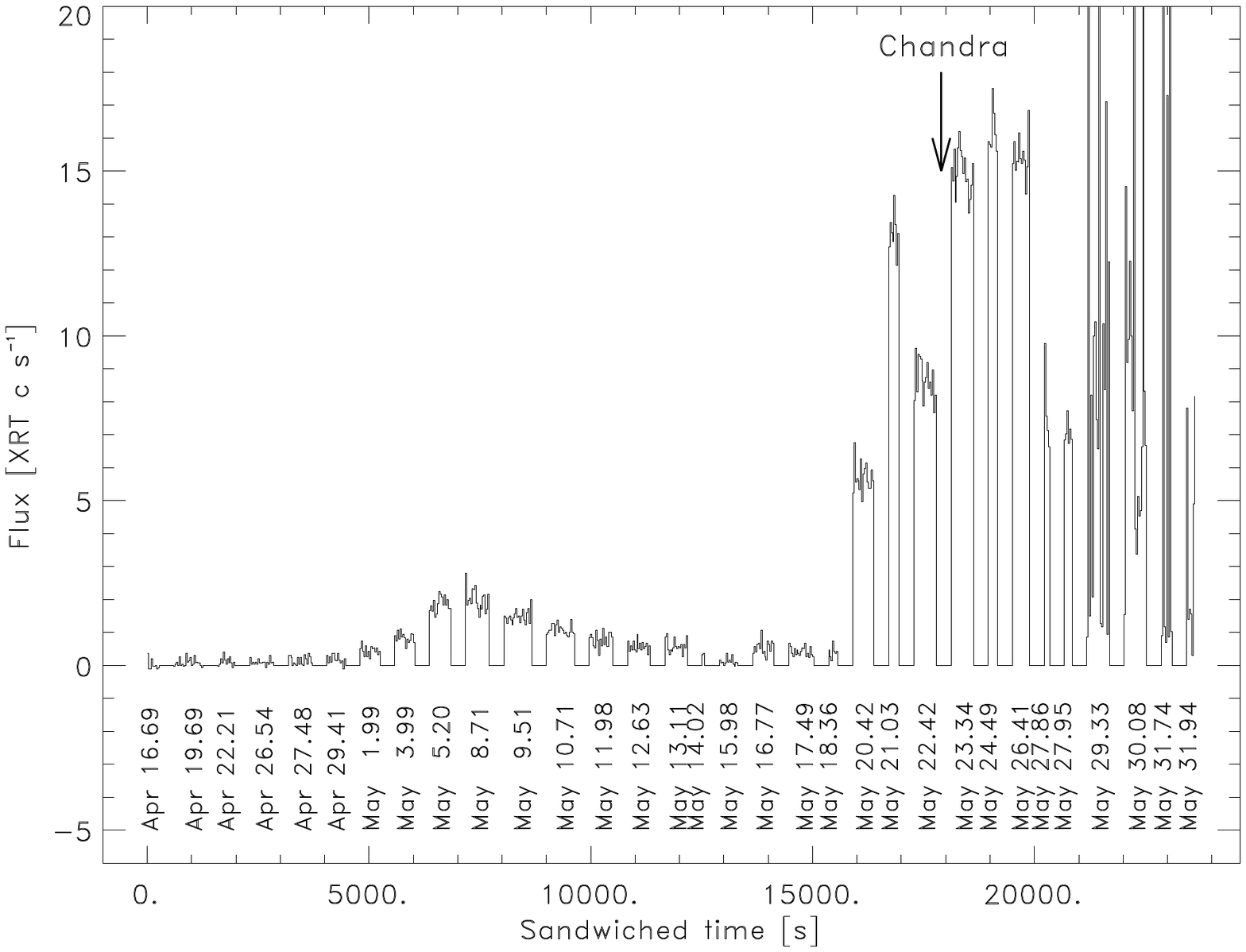}
\end{center}
\caption{Light curve of the Rapid Burster obtained with Swift-XRT. The
  time resolution is 30 s. The data have been sandwiched by largely
  eliminating the irregular data gaps between observations. Actual
  observation dates (in 2015) are displayed in the bottom part of the
  graph. For reference: May 20, 2015, is MJD 57162.  This light curve
  shows the typical evolution of a Rapid Burster outburst: first the
  banana state without type-II bursts and then, in this case on May
  29, the island state with many type-II bursts (see sandwiched light
  curve in \cite{bag15a}). The Chandra observation was in the banana
  state.\label{figxrtlc}}
\end{figure}

The transition state is particularly interesting because it represents
the onset of the peculiar long type-II bursts. It was recently
investigated through a dedicated simultaneous observation in 2015 with
Swift, XMM-Newton and NuSTAR \cite{vandeneijnden2016}. It was inferred
from a broad Fe-K emission line that the accretion disk was truncated
at 42$^{+7}_{-5}$ gravitational radii, almost 100 km away from a
canonical neutron star. This conclusion is in line with the hypothesis
that the type-II bursts are due to magnetic gating of the accretion
flow through that inner edge \cite{dangelo2012}.

The banana state, when the Rapid Burster is expected to exhibit
frequent type-I bursts because the accretion rate is high, was thus
far not covered by an observation with a high spectral
resolution. Therefore, we applied for and received a Chandra
target-of-opportunity observation of the Rapid Burster while it was in
the banana state. Here, we report our findings from that
observation. The aim was to detect as many type-I bursts as possible
and investigate their spectrum for the occurrence of absorption lines
and edges as a possible probe of the ashes of thermonuclear burning
and of the compactness of the neutron star through a measurement of
the gravitational redshift (e.g., \cite{cot02}).

\section{Observation}

\begin{figure}[t]
\begin{center}
\includegraphics[width=\columnwidth,angle=0,trim=0cm 2cm 3cm 2cm]{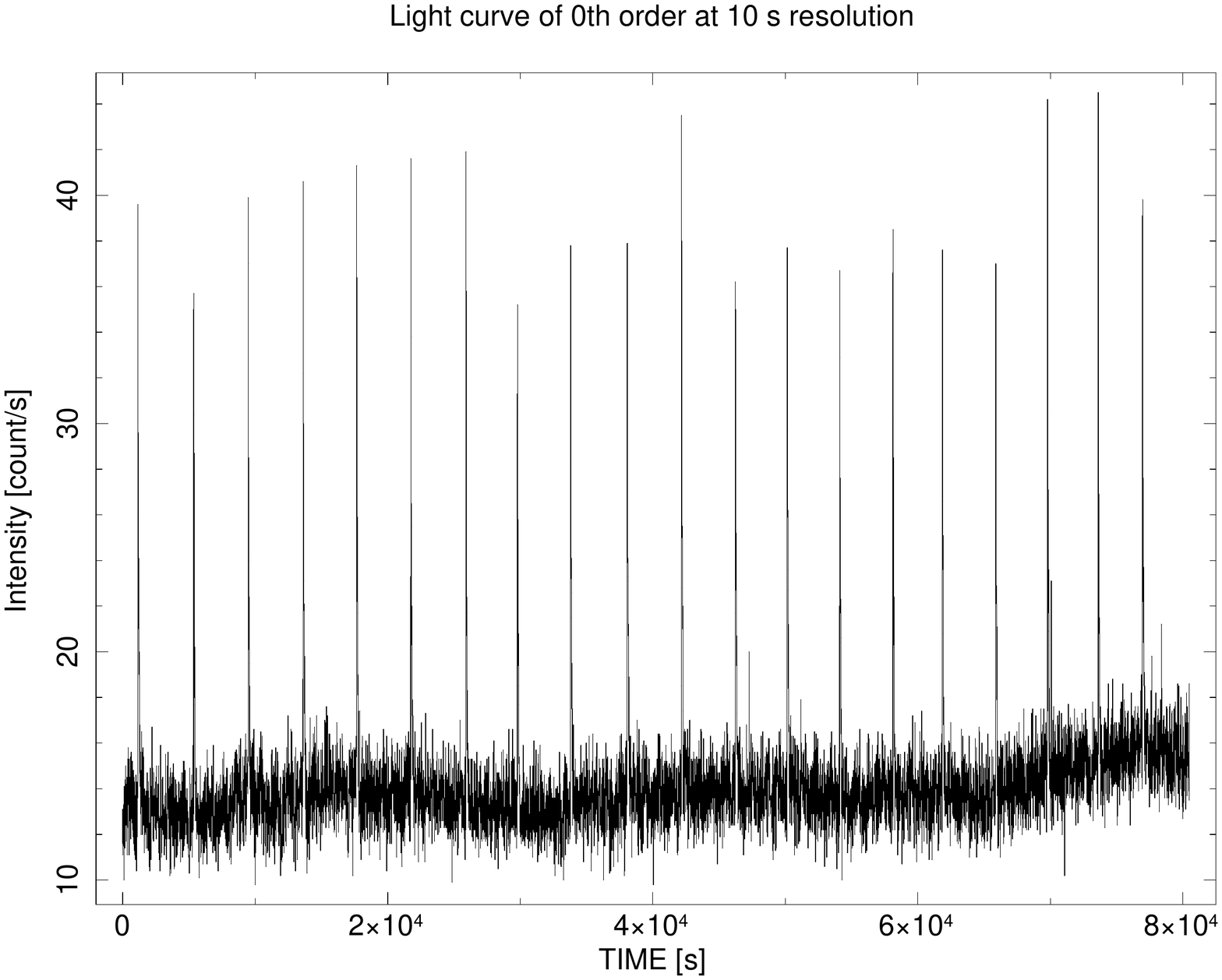}
\end{center}
\caption{Chandra light curve of all photons of order 0 at 10 s
  resolution.\label{figlc}}
\end{figure}

We started an observation campaign targeted on the outburst expected
in the spring of 2015. Since the MAXI all-sky monitor onboard the ISS
\cite{mat09} is not able to distinguish clearly between the flux from
the Rapid Burster and the 0.5~deg distant 'Slow Burster' 4U 1728-34,
we resorted to an observation campaign with the X-ray telescope XRT
onboard Swift \cite{gehrels2004}. This campaign started on April 16,
2015. The observations occurred every 3 days and lasted each about 500
s. All but one were carried out in the Windowed Timing Mode of the
XRT; the exception was in the Photon Counting Mode. The resulting
light curve is shown in Fig.~\ref{figxrtlc}. It took more than one
month before the start of the banana outburst could be unambiguously
identified, on May 20 2015 10:02 UT. There is a small precursor
outburst peaking two weeks before that. As most outbursts start with
high luminosities and the banana state, it was crucial to schedule a
follow-up observation quickly. As soon as the XRT flux reached 14 c/s
on May 21 00:39 UTC, a Chandra observation was requested and granted.

The Chandra (\cite{weisskopf2002}) observation started 1.31 d later,
on May 22, 2015, at 15:14:10 UTC (MJD 57164.63472) and ended on May 23
at 14:06:53 UTC. The exposure time is 80.19 ks. Thanks to the prompt
response, the Rapid Burster was caught while still in the banana state
and before the onset of the island state with the type-II bursts that
started five days later (c.f., Fig.~\ref{figxrtlc}).  The High-Energy
Transmission Grating Spectrometer (HETGS; \cite{canizares2005}) was
positioned into the telescope beam and ACIS-S (\cite{garmire2003})
employed as focal plane detector, enabling high-resolution
spectroscopy between 0.7 and 10 keV, although the high $N_{\rm H}$ and
high background precluded the usefulness of data below 1.5 keV. ACIS-S
was employed in the continuous clocking (CC) mode because of the high
fluxes expected from the Rapid Burster.

\section{Light curve analysis}

The light curve of all Chandra-detected photons is shown in
Fig.~\ref{figlc}. 20 type-I X-ray bursts were detected, in the longest
continuous data set with that many type-I bursts from the Rapid
Burster. The bursts occur frequently, as compared to other prolific
bursters, and regularly at intervals ranging between 4215 s (for the
first pair of bursts) and 3337 s (for the last pair) which is a change
by a factor of 1.26. This matches well the general trend of an
increasing non-burst flux level which increases by about a factor of
16~c~s$^{-1}$/13~c~s$^{-1}$=1.23, suggesting the non-burst flux to be
an accurate representation of the accretion rate.

\begin{figure}[t]
\begin{center}
\includegraphics[width=\columnwidth,angle=0,trim=0cm 2cm 3cm 2cm]{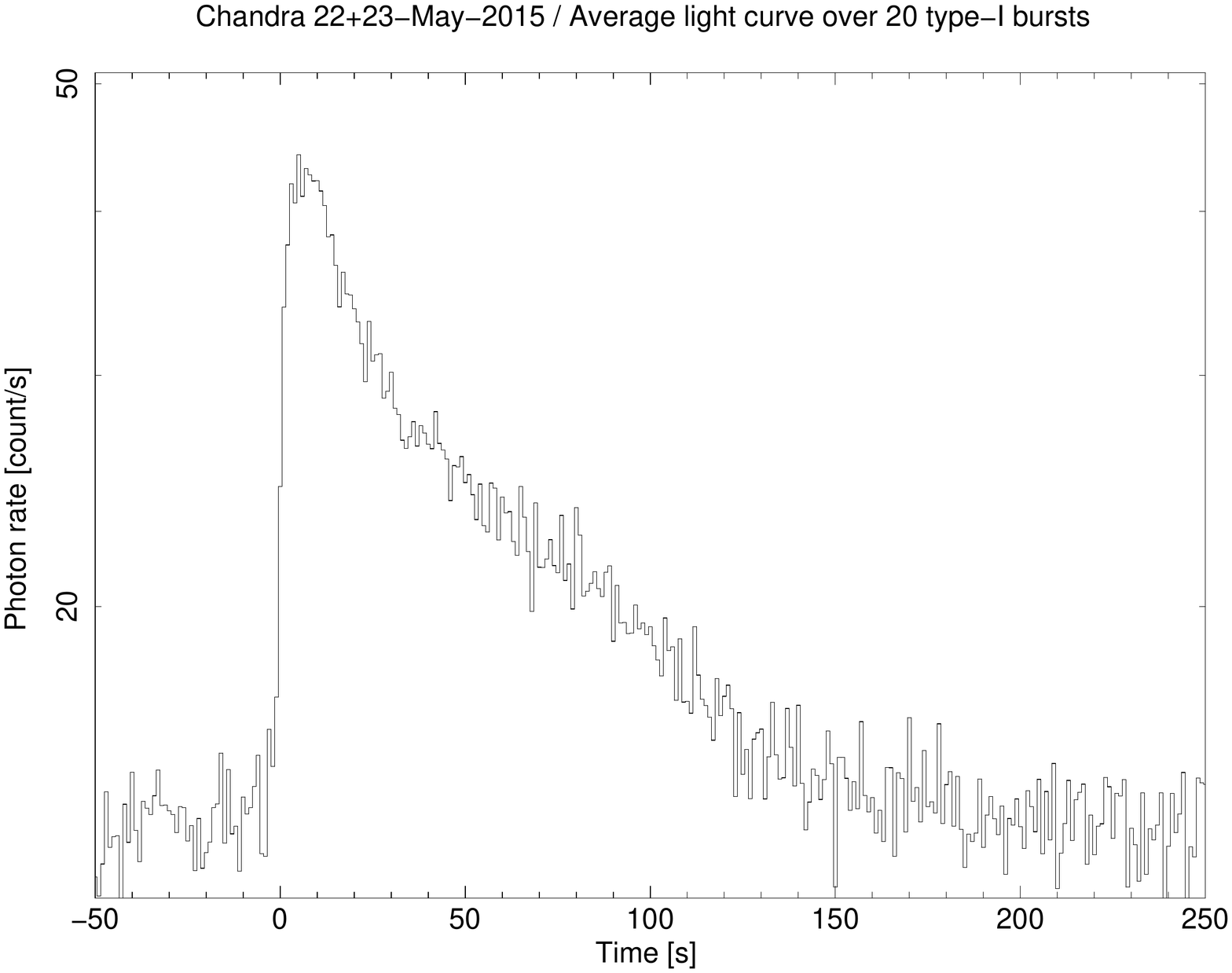}
\end{center}
\caption{Average light curve of all 20 bursts with a time resolution
  of 1 s. Note that the flux scale is logarithmic.\label{figfolded}}
\end{figure}

Figure~\ref{figfolded} shows the average light curve of all bursts
combined. Burst emission is clearly visible for about 180 s. As
typical for type-I bursts from the Rapid Burster \cite{bag13,bag15a},
the profile has a flat top of 8-10 s and a hump feature in the decay
between 30 and 120 s from the burst onset. We fitted an exponential
function to the decay part of the light curve up to 300 s after and
from 50 to 20 s before the onset, and find an e-folding decay time of
$52.3\pm2.0$~s ($\chi^2_\nu=1.29, \nu=327$). A power law function fits
the data better, provided a Gaussian contribution is included to allow
for the hump. The power law decay index is $1.21\pm0.08$
($\chi^2_\nu=1.00, \nu=324$). The Gaussian component contains
$56\pm3$\% of all burst photons. The characteristic time scale of the
burst, as determined by dividing the count fluence by the peak flux,
is 46 s.

The hump arises probably from rapid proton capture (rp) burning in a
H-rich environment. Protons are captured by a wide range of isotopes,
followed by beta decay \cite{wallace1981}. The latter is a decay
process which for some isotopes has a relatively long time scale of
tens of seconds, giving rise to the hump feature. The strength of the
rp component is quite similar to that seen in the well-studied bursts
from GS~1826-24 (e.g., \cite{galloway2004}).

\section{Spectral analysis}

\subsection{Methodology}

We employed {\tt ciao} version 4.9, CALDB version 4.7.2 (for details
of HETGS calibration, see \cite{marshall2012}) and {\tt xspec} version
12.9 for our data analysis. In particular we employed the standard
routines {\tt chandra\_repro} to reprocess the event file, {\tt
  tg\_extract} to extract for both MEG and HEG all six orders for any
particular time stretch, {\tt mktgresp} to generate (ancillary)
response files and {\tt combine\_grating\_spectra} to generate
combined spectra for all plus and minus orders. In all spectral
models, we applied absorption by the interstellar medium according to
the model by \cite{wilms2000} (model {\tt tbabs} in {\tt xspec} with
abundances fixed). Although we here report only on spectra that are
sums of negative and positive grating orders, we have inspected the
orders separately but found no additional insight from that.
We inspected all non-binned data for narrow spectral features, and as
well 2x, 4x and 8x binned data.

On the basis of the light curve at 0.1 s resolution, we divided the
80.19~ks Chandra observation into 20 burst and 21 non-burst periods.
Each burst period is 180 s (see folded burst light curve in
Fig.~\ref{figfolded}). This leaves a total non-burst exposure time of
76.6 ks.

\subsection{Non-burst spectrum}

\begin{figure}[t]
\begin{center}
\includegraphics[width=\columnwidth,angle=0,trim=0cm 0cm 3cm 0cm]{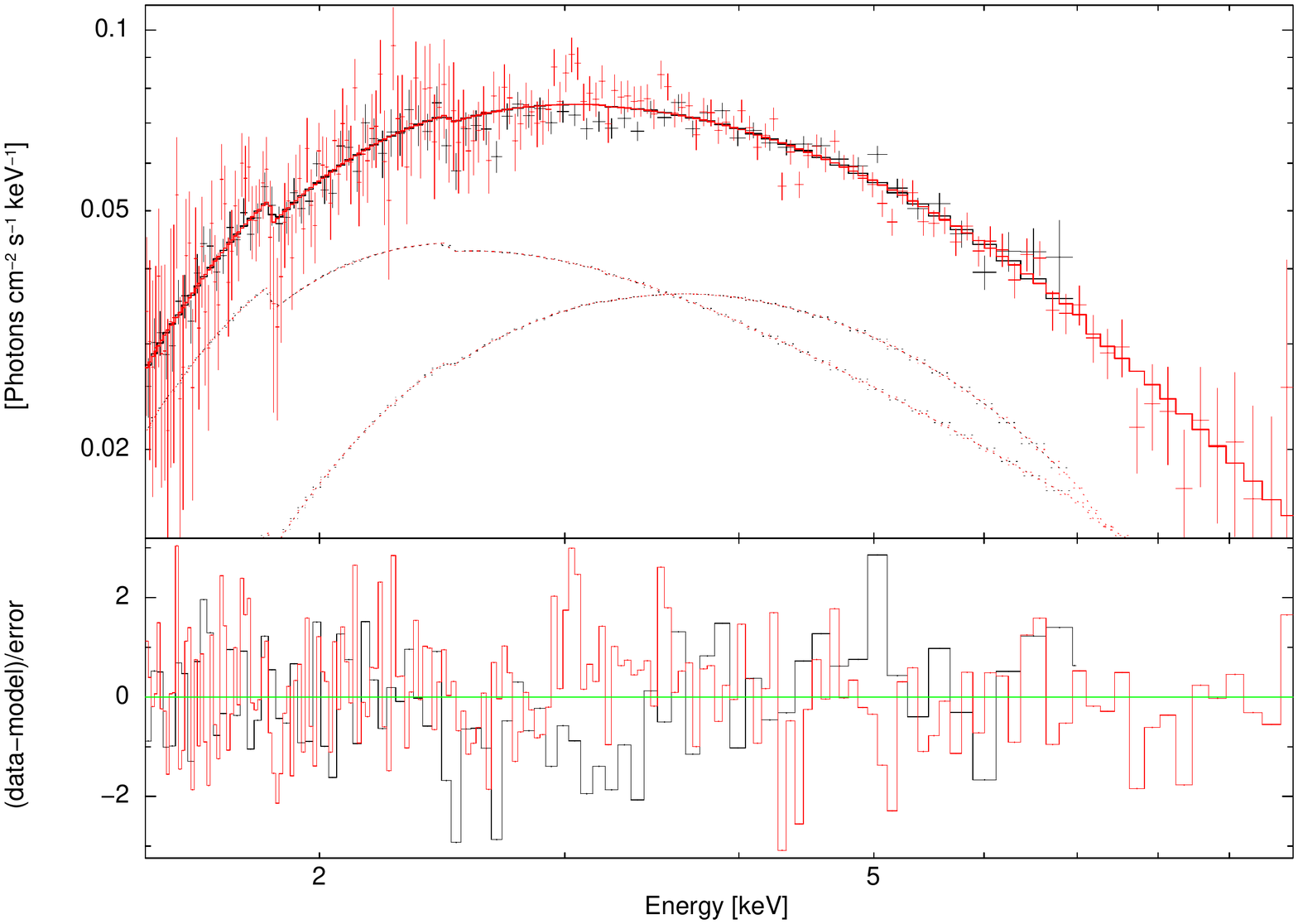}
\end{center}
\caption{Average spectrum of the first 150 s of all bursts. Black and
  red are the combined -1st and +1st order data from MEG and HEG. The
  data have been rebinned by a factor of 16.\label{figsp}}
\end{figure}

The non-burst spectrum has a photon count rate of 11.5 c~s$^{-1}$ in
all first orders of both HEG and MEG, 0.06 c~s$^{-1}$ in the second
and 1.0 c~s$^{-1}$ in the third orders. We therefore only studied
first-order data.  An absorbed power law does not fit the data well
($\chi^2_\nu$=1.967 for $\nu=4105$), but neither does an absorbed
Comptonized spectrum plus (disk) black body.  The spectra are not
consistent between the MEG and HEG with uncorrelated deviations
between the data of both instruments of up to about 10\%. The
systematic deviations between the data of HEG and MEG and the high
$\chi^2_\nu$ value are attributed to issues with the handling of CC
data (see Chandra memo by N. Schultz\footnote{\tt
  http://asc.harvard.edu/cal/Acis/ccmode/ccmode\_final\_doc03.pdf}). A
systematic uncertainty of 5.5\% per channel improves $\chi^2_\nu$ to
an acceptable value (i.e., with a chance probability of 1\%), with a
photon index of $\Gamma=1.489\pm0.008$ and $N_{\rm
  H}=(2.39\pm0.02)\times10^{22}$~cm$^{-2}$. $N_{\rm H}$ is 1.3 times
larger than the optical value (see Sect.~\ref{intro}) and 1.3 times
smaller than inferred in \cite{vandeneijnden2016}. This value is
dependent on the continuum model at low energies which is ill
determined for the Rapid Burster. The 1.5-7.0 keV flux is
$(1.462\pm0.004)\times10^{-9}$~erg~s$^{-1}$cm$^{-2}$. If we employ a
Comptonized spectrum to better approximate the bolometric flux, we
find an 1-30 keV unabsorbed flux of
$(4.3\pm0.1)\times10^{-9}$~erg~s$^{-1}$cm$^{-2}$. For the distance of
$8.1\pm1.0$ kpc, this translates to $17\pm4$\% of the
$2\times10^{38}$~erg~s$^{-1}$ Eddington limit for a H-rich
atmosphere. This is still 3 times lower than peak value of
$1.2\times10^{-8}$~erg~s$^{-1}$cm$^{-2}$ observed with RXTE
\cite{bag15b}.

We also tested the model employed on 1-30 keV XMM-Newton+NuSTAR data
by \cite{vandeneijnden2016}, consisting of a combination of a
Comptonized component, a (disk) black body, a narrow Fe-K line and a
reflection component. We find an equally good agreement with the 1.5-7
keV Chandra data as the above-mentioned simpler models when we leave
free the normalizations of all component and $N_{\rm H}$, with the
Comptonized component converging to a 3 times larger value, the black
body component 1.5 times larger, the reflection component 11 times
smaller and $N_{\rm H}=(2.78\pm0.02)\times10^{22}$~cm$^{-2}$. The
narrow Fe-K line becomes insignificant. The unabsorbed 1-30 keV flux
for this model is $2.24\times10^{-9}$~erg~s$^{-1}$cm$^{-2}$ which is
1.9 times smaller than based on the above-mentioned model which
testifies to an uncertainty of 50\% in extrapolating from 1.5-7 to
1-30 keV. The absorbed 1.5-7 keV flux of the Chandra observation is
1.5 times larger than that of the XMM-Newton/NuSTAR observation.

There is a feature at 2.4 keV which coincides with the S K-edge that
is also present in the model for the interstellar absorption. The
measured edge is deeper than expected for the fitted value of $N_{\rm
  H}$. If we leave free the S abundance and $N_{\rm H}$, we find the S
abundance to be $3.0\pm0.2$ times overabundant with respect to solar
and $N_{\rm H}=(2.35\pm0.16)\times10^{22}$~cm$^{-2}$.

We do not find other narrow spectral absorption features that are
consistent between HEG and MEG. The 3$\sigma$ upper limit on the
equivalent width for a narrow line at 6 keV (i.e., narrower than a few
eV) is 3 eV and on an absorption edge at 6 keV 60 eV. For $E=3$~keV
these upper limits are 1 eV and 30 eV, respectively. We can marginally
improve on the fit beyond 6 keV if we include either an absorption
edge at $7.1\pm0.03$ keV ($\tau=0.14\pm0.01$ and $\chi^2_\nu$ from
1.050 to 1.029 with $\nu$ from 4105 to 4103) or a relativistically
broadened Fe-K line \cite{laor1991} at $6.54\pm0.03$~keV (with a line
strength of $(4.8\pm0.4)\times10^{-3}$~phot~s~cm$^{-2}$,
$\chi^2_\nu=1.014$ with $\nu=4103$).

\subsection{Burst spectrum}

\begin{figure}[t]
\begin{center}
\includegraphics[width=\columnwidth,angle=0,trim=0cm 1cm 2cm
  20cm]{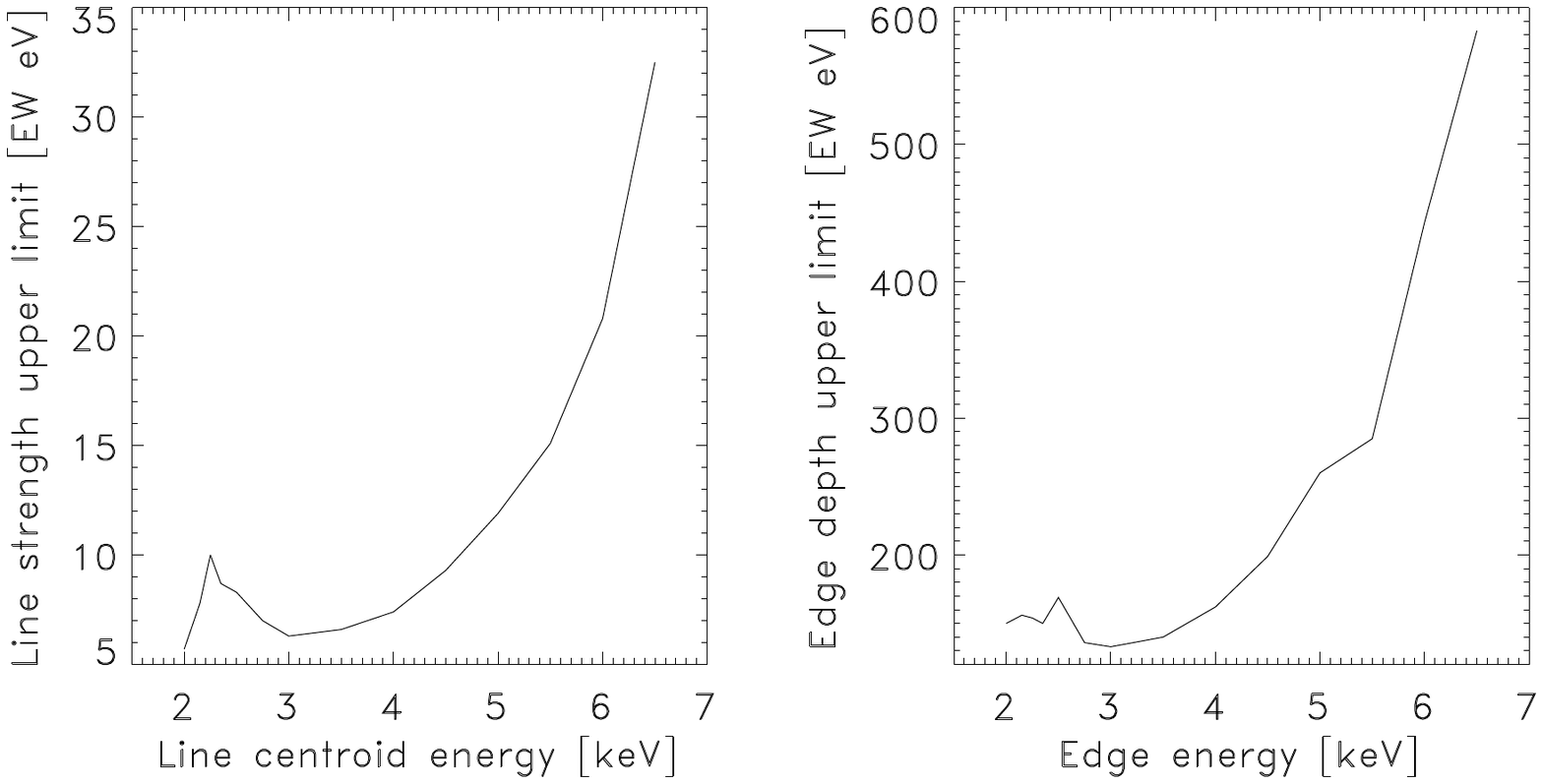}
\end{center}
\caption{Upper limit to the equivalent width of absorption lines
  (left) and absorption edges (right) as a function of line centroid
  energy and edge energy, respectively. These values are based on the
  total burst spectrum from all -1 and +1st orders of HEG and
  MEG.\label{figul}}
\end{figure}

We extracted data for the initial 150 s of all bursts (exposure time
3000 s). The average count rate of all these data is 5.8 c~s$^{-1}$
above the non-burst level. We applied a continuum model consisting of
the fixed non-burst spectrum discussed above and a black body
component, both absorbed by the same amount of $N_{\rm H}$. We
obtained an excellent continuum fit with a black body model
temperature of k$T=1.59\pm0.02$~keV ($\chi^2_\nu$=0.885 for
$\nu=4106$), see Fig.~\ref{figsp}. No significant narrow spectral
features were observed; neither were they in spectra of complete
individual bursts. It should be mentioned that we, again, found the S
K-edge at 2.4 keV to be deeper than expected for the used value of
$N_{\rm H}$, by the same amount as for the non-burst spectrum. Since
we found the same to be true for the non-burst spectrum, this edge is
not directly attributable to burst ashes.

Since there are no unambiguous detections of narrow spectral features
in the burst emission, we present in Fig.~\ref{figul} the upper
limits. For a line that is Doppler broadened by a spin of 400 Hz
\cite{bauboeck2013}, the upper limits on the line strength are 5 times
larger. For spectra of only the peak (180 s exposure time; 25.7
c~s$^{-1}$ above the non-burst level) and the initial phase of the
decay we neither find narrow spectral features with upper limits that
are three times worse for both the peak and decay spectrum. We find a
peak bolometric flux of
$(8.9\pm0.3)\times10^{-9}$~erg~s$^{-1}$cm$^{-2}$ which is consistent
with earlier studies of type-I bursts from the Rapid Burster
\cite{bag13}. The peak temperature is $1.99\pm0.06$~keV. For a
distance of 8.1 kpc, the emission area radius is $6.1\pm0.2$ km and
the peak luminosity $7\times10^{37}$~erg~s$^{-1}$. The bursts do not
reach the Eddington limit of $2\times10^{38}$~erg~s$^{-1}$ for a
canonical 1.4 M$_\odot$ neutron star with a H-rich atmosphere.

The burst $\alpha$ parameter, defined as the ratio of the fluence in
the persistent emission between bursts and that of the bursts,
is $57\pm2$. This is a typical value for H-rich bursts, in accordance
with the strong rp component \cite{lewin1993}, and consistent with
results from the Rapid Burster RXTE study \cite{bag13}.

\section{Discussion}


Weinberg et al. \cite{wei06} modeled the composition of the neutron
star photosphere in a type-I burst and predict that convection can
dredge up nuclear ashes into the photosphere when the nuclear
luminosity reaches the Eddington limit. Also, this cleans up the
atmosphere from H and He. The result can be that strong absorption
edges are imprinted into the burst spectrum. \cite{wei06} calculate
that photoionization edges of $^{60}$Zn, $^{28}$Si and $^{32}$S can
reach equivalent widths of up to 300 eV or even 900 eV for
$^{32}$S. The bursts that we investigate here do not reach the
Eddington limit though, so the predictions by \cite{wei06} can be
considered upper limits to the situation for the Rapid Burster.

The upper limits that we find for the absorption edges are of the same
order of magnitude as the predictions by \cite{wei06} for three
elements. Given the above-mentioned argument about the bursts not
being Eddington limited, it is not unexpected that we do not detect
absorption edges.

Models for absorption lines from enriched neutron star atmospheres,
including radiation transfer calculations, convection and nuclear
enrichment, are not found in the literature. Therefore, it is not
possible to make comparisons with sensible predictions and we put
forward the measured upper limits as constraints for future models.

One may wonder why it was worth doing this observation if the
predictions were close to the detection sensitivity. The Rapid Burster
may be the only neutron star spinning slowly and whose type-I bursts
can fairly easy (i.e., with a moderate amount of exposure time) be
measured with a spectrometer. Given the importance of the question
about the EOS and the lack of an unambiguous detection of narrow
spectral features sofar, this observation was considered worth the
risk.


\section{Conclusion}

We successfully acquired, through a carefully planned observation
campaign, a Chandra high-resolution grating spectrum of the Rapid
Burster while it was in the banana state.  The Chandra observation
yielded 20 type-I X-ray bursts which was the goal of this
observation. The motivation of the observation was to search for
narrow spectrum features in the burst emission. None were found. This
implies a stringent upper limit on such features for the non-burst
data of about 10 eV in equivalent width. The upper limit for
absorption edges are of the same order of magnitude as previously
obtained for SAX J1808.4-3658 (e.g., \cite{zand2013}). It should be
noted that that neutron star has a fast spin rate. The upper limits
for an absorption edge are of the same order as the predictions by
\cite{wei06} and, thus, do not unambiguously constrain that model.

The observations obtained so far with Chandra and XMM-Newton have not
been successful in finding unambiguously narrow spectral features in
the emission of bright or many combined normal type-I bursts. The only
alternative remaining with these instruments concerns the rare
intermediate duration bursts with superexpansion \cite{zan10} and
superbursts \cite{kee08b} for which such features are already strongly
suggested to exist from low spectral-resolution data (e.g.,
\cite{zan10,kajava2017}).

\vspace{5mm}\noindent {\em Acknowledgments.}  We thank the Swift and
Chandra teams for their support in making the observation campaign a
success. TB, DG and AW thank the International Space Science Institute
in Bern, Switzerland, for hosting International Teams on X-ray bursts.
We thank Jakob van den Eijnden for providing the model for the
XMM-Newton/NuSTAR spectrum in a machine readable form.


\begin{thebibliography}{99}{\parsep=0cm}
\bibitem{wei06} N.~N. {Weinberg}, L.~{Bildsten} and H.~{Schatz}, 2006,
  \emph{ApJ}, {\bf 639}, 1018--1032
\bibitem{lattimer2007} J.~M. {Lattimer} and M.~{Prakash}, 2007,
  \emph{PhysRev} {\bf 442}, 109--165
\bibitem{watts2016} A.~L. {Watts}, N.~{Andersson}, D.~{Chakrabarty},
  M.~{Feroci}, K.~{Hebeler}, G.~{Israel} et~al., 2016, \emph{Reviews of
      Modern Physics}, {\bf 88}, 021001
\bibitem{paerels2009} F.~{Paerels}, M.~{M{\'e}ndez}, M.~{Agueros},
  M.~{Baring}, D.~{Barret}, S.~{Bhattacharyya} et~al., 2009, in
  \emph{astro2010: The Astronomy and Astrophysics Decadal Survey},
  vol.~2010 of \emph{Astronomy}
\bibitem{lewin1976} W.~Lewin, J.~Doty, G.~Clark, S.~Rappaport,
  H.~Bradt, R.~Doxsey et~al., 1976, \emph{ApJ}, {\bf 207}, L95--L99
\bibitem{lewin1993} W.~H.~G. {Lewin}, J.~{van Paradijs} and
  R.~E. {Taam}, 1993, \emph{Space Science Reviews}, {\bf 62}, 223
\bibitem{saracino2015} S.~{Saracino}, E.~{Dalessandro},
  F.~R. {Ferraro}, B.~{Lanzoni}, D.~{Geisler}, F.~{Mauro} et~al., 2015
  \emph{ApJ}, {\bf 806}, 152
\bibitem{masetti2002} N.~{Masetti}, 2002, \emph{\aa}, {\bf 381},
  L45--L48
\bibitem{simon2008} V.~{Simon}, 2008, \emph{\aa}, {\bf 492}, 135--143
\bibitem{bag13} T.~{Bagnoli}, J.~J.~M. {in 't Zand}, D.~K. {Galloway}
  and A.~L. {Watts}, 2013, \emph{MNRAS}, {\bf 431}, 1947--1955
\bibitem{bag14} T.~{Bagnoli}, J.~J.~M. {in 't Zand}, A.~{Patruno} and
  A.~L. {Watts}, 2014, \emph{MNRAS}, {\bf 437}, 2790--2801
\bibitem{bag15a} T.~{Bagnoli}, J.~J.~M. {in't Zand}, C.~R. {D'Angelo}
  and D.~K. {Galloway}, 2015, \emph{MNRAS}, {\bf 449}, 268--287
\bibitem{bag15b} T.~{Bagnoli} and J.~J.~M. {in't Zand}, 2015, \emph{MNRAS}
  {\bf 450}, L52--L56
\bibitem{guerriero1999} R.~{Guerriero}, D.~W. {Fox}, J.~{Kommers},
  W.~H.~G. {Lewin}, R.~{Rutledge}, C.~B. {Moore} et~al., 1999, \emph{MNRAS},
    {\bf 307}, 179--189
\bibitem{hasinger1989} G.~{Hasinger} and M.~{van der Klis}, 1989,
  \emph{\aa}, {\bf 225}, 79--96
\bibitem{patruno2012} A.~{Patruno} and A.~L. {Watts}, 2012,
  arxiv.org:1206.2727
\bibitem{vandeneijnden2016} J.~{van den Eijnden}, T.~{Bagnoli},
  N.~{Degenaar}, A.~M. {Lohfink}, M.~L.  {Parker}, J.~J.~M. {in 't
    Zand}, A.C.~{Fabian}, 2017, \emph{MNRAS}, {\bf 466}, L98-102
\bibitem{dangelo2012} C.~R. {D'Angelo} and H.~C. {Spruit}, 2012,
  \emph{MNRAS}, {\bf 420}, 416--429
\bibitem{cot02} J.~{Cottam}, F.~{Paerels} and M.~{Mendez}, 2002,
  \emph{Nat}, {\bf 420}, 51--54
\bibitem{mat09} M.~{Matsuoka}, K.~{Kawasaki}, S.~{Ueno}, H.~{Tomida},
  M.~{Kohama}, M.~{Suzuki} et~al., 2009, \emph{PASJ}, {\bf 61}, 999
\bibitem{gehrels2004} N.~{Gehrels}, G.~{Chincarini}, P.~{Giommi},
  K.~O. {Mason}, J.~A. {Nousek}, A.~A. {Wells} et~al., 2004,
  \emph{ApJ}, {\bf 611}, 1005--1020
\bibitem{weisskopf2002} M.~C. {Weisskopf}, B.~{Brinkman},
  C.~{Canizares}, G.~{Garmire}, S.~{Murray} and L.~P. {Van
    Speybroeck}, 2002, \emph{PASP}, {\bf 114}, 1--24
\bibitem{canizares2005} C.~R. {Canizares}, J.~E. {Davis}, D.~{Dewey},
  K.~A. {Flanagan}, E.~B. {Galton}, D.~P. {Huenemoerder} et~al., 2005,
  \emph{PASP}, {\bf 117}, 1144--1171
\bibitem{garmire2003} G.~P. {Garmire}, M.~W. {Bautz}, P.~G. {Ford},
  J.~A. {Nousek} and G.~R.  {Ricker}, Jr., 2003, in \emph{X-Ray and
    Gamma-Ray Telescopes and Instruments for Astronomy.}
  (J.~E. {Truemper} and H.~D. {Tananbaum}, eds.), vol.~4851 of
  \emph{Proc. SPIE}, pp.~28--44
\bibitem{wallace1981} R.~K. {Wallace} and S.~E. {Woosley}, 1981,
  \emph{ApJs}, {\bf 45}, 389--420
\bibitem{galloway2004} D.~K. {Galloway}, A.~{Cumming}, E.~{Kuulkers},
  L.~{Bildsten}, D.~{Chakrabarty} and R.~E. {Rothschild}, 2004, \emph{ApJ},
    {\bf 601}, 466--473
\bibitem{marshall2012} H.~L. {Marshall}, 2012, in \emph{Space
  Telescopes and Instrumentation 2012: Ultraviolet to Gamma Ray},
  vol.~8443 of \emph{Proc. SPIE}, p.~844348
\bibitem{wilms2000} J.~{Wilms}, A.~{Allen} and R.~{McCray}, 2000,
  \emph{ApJ}, {\bf 542}, 914--924
\bibitem{laor1991} A.~{Laor}, 1991, \emph{ApJ},  {\bf 376}, 90--94
\bibitem{bauboeck2013} M.~{Baub{\"o}ck}, D.~{Psaltis} and
  F.~{{\"O}zel}, 2013, \emph{ApJ}, {\bf 766}, 87
\bibitem{zand2013} J.~J.~M. {in 't Zand}, D.~K. {Galloway},
  H.~L. {Marshall}, D.~R. {Ballantyne}, P.~G. {Jonker},
  F.~B.~S. {Paerels} et~al., 2013, \emph{A\&A} {\bf 553} A83
\bibitem{zan10} J.~J.~M. {in 't Zand} and N.~N. {Weinberg}, 2010,
  \emph{\aa}, {\bf 520}, A81
\bibitem{kee08b} L.~{Keek} and J.~J.~M. {in 't Zand}, 2008, in
  \emph{Proceedings of the 7th INTEGRAL Workshop}, Proc. of Sc.,
  arxiv.org:0811.4574
\bibitem{kajava2017} J.~J.~E. {Kajava}, J.~{N{\"a}ttil{\"a}},
  J.~{Poutanen}, A.~{Cumming}, V.~{Suleimanov} and E.~{Kuulkers},
  2017, \emph{MNRAS}, {\bf 464}, L6--L10
\end{thebibliography}

\end{document}